\documentclass[conference]{IEEEtran}
\IEEEoverridecommandlockouts
\usepackage{cite}
\usepackage{amsmath,amssymb,amsfonts}
\usepackage{braket}
\usepackage{algorithmic}
\usepackage{graphicx}
\usepackage{textcomp}
\usepackage{xcolor}
\usepackage{siunitx} 
\usepackage{booktabs}  
\usepackage{hyperref}
\def\BibTeX{{\rm B\kern-.05em{\sc i\kern-.025em b}\kern-.08em
    T\kern-.1667em\lower.7ex\hbox{E}\kern-.125emX}}

\begin{document}
\title{Noise-Mitigated Variational Quantum Eigensolver with Pre-training and Zero-Noise Extrapolation
}

\author{\IEEEauthorblockN{1\textsuperscript{st} Wanqi Sun, 3\textsuperscript{rd} Chenghua Duan}
\IEEEauthorblockA{\textit{School of Electronic, Electrical and Communication Engineering} \\
\textit{University of Chinese Academy of Sciences}\\
Beijing, China \\
sunwanqi19@mails.ucas.ac.cn, cduan@ucas.ac.cn}
\and
\IEEEauthorblockN{2\textsuperscript{nd} Jungang Xu\textsuperscript{*}}
\IEEEauthorblockA{\textit{School of Computer Science and Technology} \\
\textit{University of Chinese Academy of Sciences}\\
Beijing, China \\
xujg@ucas.ac.cn}
}

\maketitle

\begin{abstract}
As a hybrid quantum-classical algorithm, the variational quantum eigensolver is widely applied in quantum chemistry simulations, especially in computing the electronic structure of complex molecular systems. However, on existing noisy intermediate-scale quantum devices, some factors such as quantum decoherence, measurement errors, and gate operation imprecisions are unavoidable. To overcome these challenges, this study proposes an efficient noise-mitigating variational quantum eigensolver for accurate computation of molecular ground state energies in noisy environments. We design the quantum circuit with reference to the structure of matrix product states and utilize it to pre-train the circuit parameters, which ensures circuit stability and mitigates fluctuations caused by initialization. We also employ zero-noise extrapolation to mitigate quantum noise and combine it with neural networks to improve the accuracy of the noise-fitting function, which significantly eliminates noise interference. Furthermore, we implement an intelligent grouping strategy for measuring Hamiltonian Pauli strings, which not only reduces measurement errors but also improves sampling efficiency. 
We perform numerical simulations to solve the ground state energy of the $H_4$ molecule by using MindSpore Quantum framework, and the results demonstrate that our algorithm can constrain noise errors within the range of $\mathcal{O}(10^{-2}) \sim \mathcal{O}(10^{-1})$, outperforming mainstream variational quantum eigensolvers. This work provides a new strategy for high-precision quantum chemistry calculations on near-term noisy quantum hardware. 
\end{abstract}


\section{Introduction}
Conducting quantum chemistry simulations on high-performance classical computers has become a crucial method for investigating the physical and chemical properties of materials. However, accurately solving the Schrödinger equation involves exponential complexity, severely limiting the size of chemical systems that can be simulated. Recent advancements in quantum computing offer a feasible solution to this problem, with the potential to achieve high-precision solutions to the Schrödinger equation with polynomial complexity on quantum computers \cite{cao2019quantum, mcardle2020quantum}.

In this context, the Variational Quantum Eigensolver (VQE), as a hybrid quantum-classical algorithm, has become an important tool for achieving this goal \cite{tilly2022variational,cerezo2022variational}. Matrix product states (MPS) are also widely applied as the VQE because their one-dimensional chain structure is effective in capturing the local features of quantum states \cite{xu2024mps-vqe,guo2023differentiable,javanmard2024matrix}.
The primary objective of the VQE is to determine the ground state energy and corresponding quantum state of the Hamiltonian $H$ for a closed quantum system, which is crucial for predicting various chemical properties including reaction rates and stable molecular configurations \cite{helgaker2013molecular,moll2018quantum,bartlett1989alternative}.
In 2014, Peruzzo et al. first applied the Variational Quantum Eigensolver (VQE) combined with unitary coupled cluster theory to quantum chemistry simulations \cite{peruzzo2014variational}, successfully determining the ground state energy of the He-H\(^+\) system.

Current Noisy Intermediate-Scale Quantum (NISQ) devices are constrained by inherent noise from physical implementations, making the accurate determination of ground state energy a challenging task. 
To address this, researchers utilize mutual information and classical algorithms to reduce entangling operations \cite{zhang2021mutual} and explore other hardware-efficient ansatz \cite{kandala2017hardware,rattew2019domain,tang2021qubit,zhang2022variational} to reduce the number of gates in the circuit. 
There has emerged a significant amount of research on adaptive ansatz construction recently \cite{grimsley2019adaptive,liu2021efficient,feniou2023overlap,ramoa2024reducing}. These approaches focus on optimizing quantum circuit design to decrease circuit depth and operational complexity, which in turn reduces noise and errors.

To tackle the noise challenges in current quantum devices, we propose an innovative variational quantum eigensolver. The main contributions of our work are as follows: 
\begin{enumerate}    
    \item Design one adjustable quantum circuit structure based on MPS, fully utilizing the chain structure of MPS to capture local entanglement, while considering a brick-wall structure with similar error scaling to ensure the circuit depth remains shallow.
    \item Pre-train quantum circuit parameters with MPS on classical computers to effectively avoid the interference of initialization on the stability of quantum circuit optimization.     
    \item Employ zero-noise extrapolation (ZNE) techniques for noise error mitigation and combine it with neural network for noisy data fitting to ensure the ground state energy solving precision.  
    \item Implement grouping measurements of Hamiltonian Pauli strings, which reduces the number of sampling and mitigates measurement noise.  
\end{enumerate}  

The numerical simulations are mainly performed by using MindSpore Quantum \cite{mindspore_quantum} 0.9.11 framework, where the noise environment includes depolarizing noise, thermal relaxation noise, and bit-flip noise. The experimental results demonstrate that our algorithm can limit noise errors within $\mathcal{O}(10^{-2}) \sim \mathcal{O}(10^{-1})$, surpassing mainstream variational eigensolvers.

\begin{figure*}[tp]
\includegraphics[width=1\textwidth]{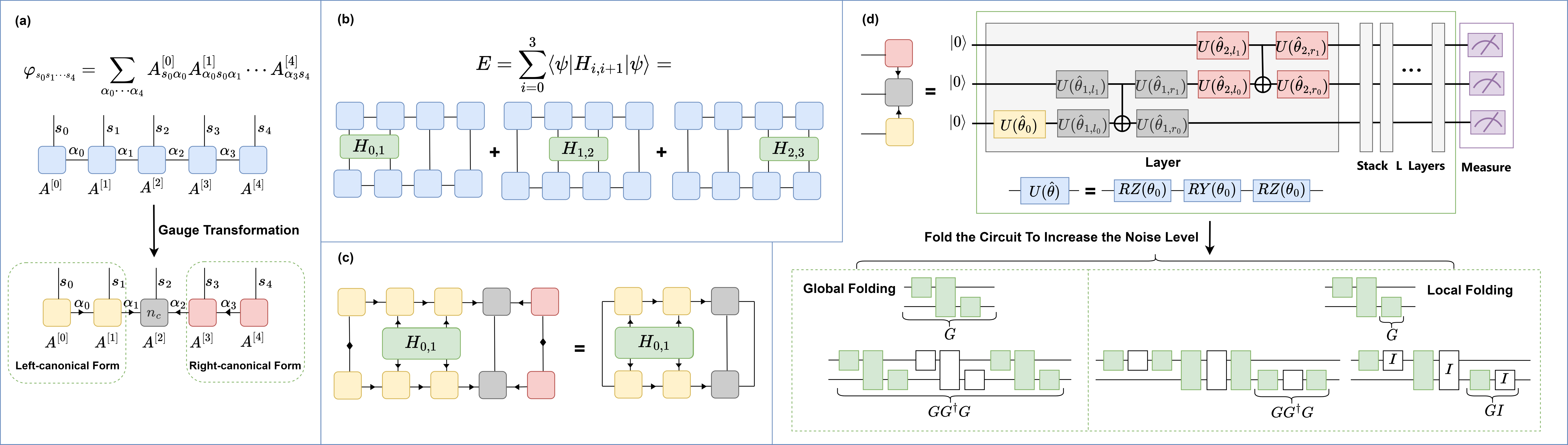} 
\caption{\label{MPS_DMRG} Schematic Representation of MPS and the Algorithm. (a) depicts a 5-qubit quantum state $\ket{\psi}$ in the MPS form, where the $n$th local tensor $A^{[n]}$ corresponds to the $n$th qubit. The MPS shown below has been gauged into a center-orthogonal form with $n_c$ as the orthogonal center. 
(b) describes the eigenvalue computation process when the MPS is employed as the trial state.
(c) illustrates how the center-orthogonal MPS can reduce the computational cost of eigenvalue calculations, requiring only the contraction of tensors from the $i$th to the $n_c$th site with the Hamiltonian coefficient tensor. 
(d) outlines our MPS-VQE algorithm. The pre-trained MPS parameters serve as the initial values for the quantum circuit parameters $\theta$, which are then further optimized. The circuit folding is employed to obtain expectation values with different noise levels, and then fits the values to extrapolate the noise-free expectation value.
}
\end{figure*}
\section{Method}
\subsection{Variational Quantum Eigensolver Framework}
The VQE is a hybrid quantum-classical algorithm designed to compute the ground state energy of quantum systems. The main implementation of VQE involves preparing a parameterized trial wavefunction $|\psi(\boldsymbol{\theta})\rangle$ on a quantum device, then combining it with classical machine learning optimization algorithms such as gradient descent to continuously adjust the parameters $\boldsymbol{\theta}$ until the expectation value $\langle\psi (\boldsymbol {\theta}) | H| \psi(\boldsymbol{\theta})\rangle$ is minimized. Mathematically, this is expressed as:
\begin{equation}
E_0=\min_{\boldsymbol{\theta}}\langle\psi(\boldsymbol{\theta})|H|\psi(\boldsymbol{\theta})\rangle, \label{E_min}
\end{equation}
where $E_0$ denotes the ground state energy of the system. 
In quantum chemistry simulations, the Hamiltonian is typically expressed as a linear combination of a set of Pauli operators. For a system of \( n \) qubits, the Hamiltonian can be formally represented as
\begin{equation}
    H=\sum_{i=1}^Mc_iH_{i}
\end{equation}
where \( c_i \) is real coefficient and \( H_i \) is the tensor product of Pauli operators, such as \( H_i=\sigma_{i_1}\otimes\sigma_{i_2}\otimes\cdots\otimes\sigma_{i_n} \),
$\sigma_{i_k}\in\{I,X,Y,Z\}.$

The procedure of proposed variational quantum eigensolver algorithm is as follows:
\begin{enumerate}  
    \item \textit{Preparation of Initial State}: Initialize the quantum state as Hartree-Fock state.
    \item \textit{Wavefunction Construction}: Construct the wavefunction in the matrix product state form.  
    \item \textit{Wavefunction Preprocessing}: Gauge the wavefunction $\ket{\psi}$ to the center-orthogonal form to enhance computational stability and reduce the complexity of the expectation computation.    
    \item \textit{Quantum Circuit Design for the Wavefunction}: Design a hardware-efficient parameterized quantum circuit to prepare the preprocessed wave functions.  
    \item \textit{Pre-training}: Pre-train the matrix product state on classical computer as the quantum circuit initialization.   
    \item \textit{Quantum State Measurement}: Combine neural networks with zero-noise extrapolation techniques to mitigate measurement errors.  
         Perform grouping measurements of the Hamiltonian Pauli strings to reduce the number of measurements required.  
    \item \textit{Classical Optimization}: Compute the current energy $E(\theta)$ of the quantum system on a classical computer and use the Stochastic Gradient Descent (SGD) optimizer to update the circuit parameters.    
    \item \textit{Circuit Update}: Apply the new parameters to the quantum circuit.  
    \item \textit{Iterative Optimization}: Repeat steps 6-8 until convergence.  
\end{enumerate}

\subsection{Wavefunction Construction}
We construct the parameterized trial wavefunction \(|\psi(\boldsymbol {\theta}) \rangle\) according to the Matrix Product States (MPS), the structure of which is illustrated in Figure \ref{MPS_DMRG} (a). The MPS efficiently decomposes the coefficient global tensor \(\varphi_{s_0s_1\cdots s_{N-1}}\) of the quantum state into a product of local tensors \(\{A^{\left[n\right]}\}\). Mathematically, for a quantum many-body system with \(N\) qubits, its global quantum state \(|\psi\rangle\) can be expressed as:
\begin{equation}
	|\psi\rangle = \sum_{s_0,s_1,\cdots,s_{N-1}} \varphi_{s_0s_1\cdots s_{N-1}} \bigotimes_{n=0}^{N-1} |s_n\rangle  , \label{quantum MPS}	
\end{equation}
where
\begin{equation}
    \small  
    \varphi_{\scriptscriptstyle s_0 s_1\cdots s_{N-1}} 
    = \sum_{\scriptscriptstyle \alpha_0,\cdots,\alpha_{N-2}} A^{\left[0\right]}_{\scriptscriptstyle s_0,\alpha_0} A^{\left[1\right]}_{\scriptscriptstyle s_1,\alpha_0,\alpha_1} \cdots A^{\left[N-1\right]}_{\scriptscriptstyle s_{N-1},\alpha_{N-2}}.
\end{equation}
The \(\{\left|s_n\right\rangle\}\) denotes a set of orthonormal computational bases.
As shown in Figure \ref{MPS_DMRG} (a), $A^{\left[n\right]}_{s_n,\alpha_{n-1},\alpha_n}$ is the local tensor corresponding to the $n$-th qubit, where $\{s_n\}$ is a set of the physical indices of dimension $d$, and $\{\alpha_n\}$ is a set of the virtual indices of dimension $\chi$.

Now, the eigenvalue problem has been transformed from Eq. (\ref{E_min}) to $E_0 = \min_{\boldsymbol{\theta}} \sum_i c_i \langle\psi|H_{i,i+1}|\psi\rangle$, where the parameter $\boldsymbol{\theta}$ refers to $\{A^{[n]}\}$.

\subsection{Wavefunction Preprocessing} 
It is necessary to gauge the wavefunction $\ket{\psi}$ to the center-orthogonal form as shown in Figure \ref{MPS_DMRG} (a). Define $n_c$ as the orthogonal center. The tensor on the left side of the orthogonal centre satisfies the left orthogonality condition $\sum_{s_{n}\alpha_{n-1}}A_{\alpha_{n-1}s_{n}\alpha_{n}}^{n}A_{\alpha_{n-1}s_{n}\alpha_{n}^{\prime}}^{n *}=I_{\alpha_{n}\alpha_{n}^{\prime}}(1\leqslant n\leqslant n_{c}-1)$, and the tensor on the right side satisfies the right orthogonality condition $\sum_{s_n\alpha_n}A_{\alpha_{n-1}s_n\alpha_n}^nA_{\alpha_{n-1}^{\prime}s_n\alpha_n}^{n *}=I_{\alpha_{n-1}\alpha_{n-1}^{\prime}}(n_c+1\leqslant n\leqslant N-1)$.


One of the advantages of orthogonalization is greater stability in updating the local tensor. Only the tensor at the orthogonal center is updated at a time, and the other tensors are treated as known tensors.
The orthogonal center is moved from the leftmost to the rightmost and then from the rightmost back to the leftmost, thus updating each local tensor in the MPS twice. Such a complete left-to-right and then right-to-left update cycle is called a sweep.

Another key advantage is that it can significantly simplify the computation of the energy. 
As shown in Fig. 1(b), if the local coupling is located to the left of the orthogonal center, i.e., $i < n_c$, according to the left-right orthogonality condition, the contraction of the tensor to the left of the $i$th tensor is equal to the identity, and the contraction of the tensor to the right of the orthogonal center $n_c$ is also equal to the identity.
It is only required to contract the $i$th tensor to $n_c$th one with the Hamiltonian coefficient tensor. The simplification of the computation is exactly similar when $ i+1 \geq n_c$.

\subsection{Wavefunction Pre-training}
It is found that even for the same quantum circuit structure, if the initial parameter values are different, the measurement results of the trained quantum circuit will be different. Therefore, we adopted the MPS pre-training to initialize the quantum circuit parameters, which can preserve the local features of the quantum state in advance and improve the accuracy of the final calculation.
The objective function for MPS pre-training is defined in Eq. (\ref{E_min}) and the update rule for the $n$th local tensor is defined as follows.
\begin{equation}
    A^{[n]} \leftarrow A^{[n]} - \eta \frac{\partial \sum_i \bra{\psi} H_{i,i+1} \ket{\psi}}{\partial A^{[n]}}
\end{equation}
where $\eta$ denotes the learning rate.
\subsection{Designing Quantum Circuits for the Wavefunction}
In this paper, with reference to the structure of MPS for the circuit framework of wave functions, the $n$th local tensor $A^{[n]}$ simulates the quantum state evolution and entanglement on the $n$th qubit. The quantum circuit framework is shown in Figure \ref{MPS_DMRG} (c), where the local tensor \( A^{[n]}\) is mapped to a parameterized quantum circuit \( 
 \mathcal{U}^{[n]\dagger} \). When constructing the wave function \(|\psi\rangle\), the operators \(\{\mathcal{U}^{[n]\dagger}\}\) act sequentially on the initial Hartree-Fock state \(|\psi_{\mathrm{HF}}\rangle\):
\begin{equation}
|\psi\rangle =\mathcal{U}^{[0]\dagger}\mathcal{U}^{[1]\dagger}\cdots \mathcal{U}^{[N-1]\dagger}|\psi_{\mathrm{HF}}\rangle. \label{MPS_PQC_MAP}
\end{equation}
For $n > 0$, the 2-qubit unitary $\mathcal{U}^{[n]}$ is factorized as:
\begin{equation}  
\small  
\mathcal{U}^{[n]\dagger} = (U(\theta_{n,l_0}) \otimes U(\theta_{n,l_1})) \  CNOT \ (U(\theta_{n,r_0}) \otimes U(\theta_{n,r_1})),  
\end{equation}
where $U(\theta_{*})$ is a single qubit unitary that can be further decomposed as: 
\begin{equation}
    U(\theta_{*})=RZ(\theta_{0})RY(\theta_{1})RZ(\theta_{2}). \label{RZYZ}
\end{equation}
For $n=0$, $\mathcal{U}^{[0]}$ is a single qubit unitary, which is expressed as Eq. (\ref{RZYZ}).

This decomposition ensures that each qubit ends in a Z-axis rotation at the end of the circuit. The Pauli Z-gate has eigenvalues of 0 and 1, and its eigenvectors are consistent with the computational basis. Therefore, the energy value can be quickly calculated directly by sampling the measured quantum state, which improves the computational efficiency.

\subsection{Noise Error Mitigation}
When executing algorithms on quantum computing devices, noise interference is inevitable. For gate operation noise and relaxation noise, we apply the zero-noise extrapolation technique to mitigate the errors at the algorithmic level. This process includes two steps:\\
\textit{Step 1: Intentionally scale noise.} A technique to increase the noise level of a circuit at the gate level is to increase its depth. This can be obtained using either unitary folding or identity scaling as shown in Figure \ref{MPS_DMRG} (c).
We perform the mapping \(G \mapsto G G^\dagger G\) or $GI$ to achieve circuit folding. \\
\textit{Step 2: Extrapolate to the noiseless limit.} This is performed by fitting a curve (often called extrapolation model) to the measured expectation values at different noise levels to extrapolate the noiseless expectation values. Extrapolation can be done in practice as follows:
\begin{enumerate}
    \item Choose an extrapolation model: suppose that the expectation \(E(\lambda)\) can be described by some function \(f(\lambda; p_1, p_2, \dots, p_m)\), where \(f\) is an extrapolation model that depends on the noise-scaling parameter \(\lambda\) and a set of real parameters \(p_1, p_2, \dots, p_m\). Common extrapolation models include linear, polynomial and exponential models. In our work, a simple neural network consisting of three fully connected layers is used to fit the expected value function under different noise levels. The fitted model trained using the neural network is more flexible and the extrapolated expectation values for different noise environments are more accurate.
     \item Fit data: the extrapolation model \(f\) is used to fit different noise scaling expectations measured to obtain a set of best-fit parameters \(\tilde p_1, \tilde p_2, \dots, \tilde p_m\).
    \item Extrapolation of the noiseless limit: the expectation value for the noiseless case is obtained by computing \(f(0; \tilde p_1, \tilde p_2, \dots, \tilde p_m)\).
\end{enumerate}

\begin{table}[th]   
    \begin{minipage}[t]{0.55\textwidth}  
        \centering  
        \caption{Comparison of circuit metrics for different VQE models}  
        \label{vqecircuits}  
        \begin{tabular}{lccccc}  
            \toprule  
            \ & MPS-VQE & UCCSD & HE-VQE & Qubit UCC & SE Ansatz \\
            \midrule  
             Qubits & 8 & 8 & 8 & 8 & 8 \\
            Total gates & 91 & 2688 & 55 & 888 & 64 \\
            Parameter gates & 84 & 160 & 48 & 104 & 32 \\
            Barriers & 0 & 640 & 0 & 0 & 0 \\
            \bottomrule  
        \end{tabular}  
    \end{minipage}  
    
    \begin{minipage}[t]{0.55\textwidth}  
        \centering  
        \caption{Comparison of VQE model performance in noiseless and noisy environments for the ground state energy of the H\textsubscript{4} molecule}  
        \label{results}  
        \begin{tabular}{c c c}  
            \hline  
            Model & Noiseless (Hartree) & Noisy (Hartree) \\
            \hline  
            $\mathbf{MPS-VQE}$ & $\mathbf{-2.1609}$ & $\mathbf{-2.1490}$ \\
            HE-VQE\cite{kandala2017hardware} & -2.1723 & -1.6726 \\
            Qubit UCC\cite{yordanov2020efficient} & -2.1476 & -0.6916 \\
            SE Ansatz\cite{schuld2020circuit} & -2.1200 & -1.5781 \\
            UCCSD\cite{peruzzo2014variational} & -2.1615 & -0.5293 \\ \hline  
            FCI Benchmark & -2.1664 & -2.1664 \\
        \end{tabular}  
    \end{minipage}  
\end{table}

Since measurements of quantum states are probabilistic, multiple measurements are required to obtain the expectation value. Since state reset and re-evolution are required for each measurement, in order to improve the efficiency of measurement, the Pauli terms can be reasonably grouped so as to obtain the expectation values of multiple terms simultaneously in a single measurement. 
If the two Pauli terms \(H_i\) and \(H_j\) satisfy the commutation relation, i.e., \([H_i, H_j] = H_i H_j - H_j H_i = 0\), they share the same set of eigenstates and can be measured simultaneously in a single measurement.
Therefore, we can simultaneously measure the commuting Hamiltonians to reduce the number of measurements, thereby lowering the measurement error.

\section{Experimental Results}
\textit{Experimental Setup.} In this work, we select the hydrogen tetramer $H_4$ as the study object, with the molecular configuration \texttt{['H 0 0 1 Å', 'H 0 0 2 Å', 'H 0 0 3 Å', 'H 0 0 4 Å']}, which is an assumed model and may not represent the true molecular structure. We employ the minimal STO-3G basis set for the calculations.  

The numerical simulations are mainly performed by using MindSpore Quantum \cite{mindspore_quantum} 0.9.11 framework, and the simulator is set to \texttt{'mqvector'}. The updated code is available at \cite{code_link}. We define a specific noise model: single-qubit gates are accompanied by a single-qubit depolarizing channel noise with a polarization rate of 0.001 and thermal relaxation noise ($T_1 = \SI{100}{\micro\second}$, $T_2 = \SI{50}{\micro\second}$, $t_\text{gate} = \SI{30}{\nano\second}$); two-qubit gates are accompanied by a two-qubit depolarizing channel noise with a polarization rate of 0.004 and thermal relaxation noise ($T_1 = \SI{100}{\micro\second}$, $T_2 = \SI{50}{\micro\second}$, $t_\text{gate} = \SI{80}{\nano\second}$); measurement gates are accompanied by a bit-flip channel noise with a flip probability of 0.05.  

Table \ref{vqecircuits} summarizes our MPS-VQE circuit implementation and other models, including key metrics such as the number of qubits, total gates, barriers, and parameterized gates. The number of layers in our MPS-VQE circuit is set to 1.

\begin{figure}[tp]
 \centering
    \begin{minipage}[t]{0.4\textwidth}  
    \includegraphics[width=\textwidth]{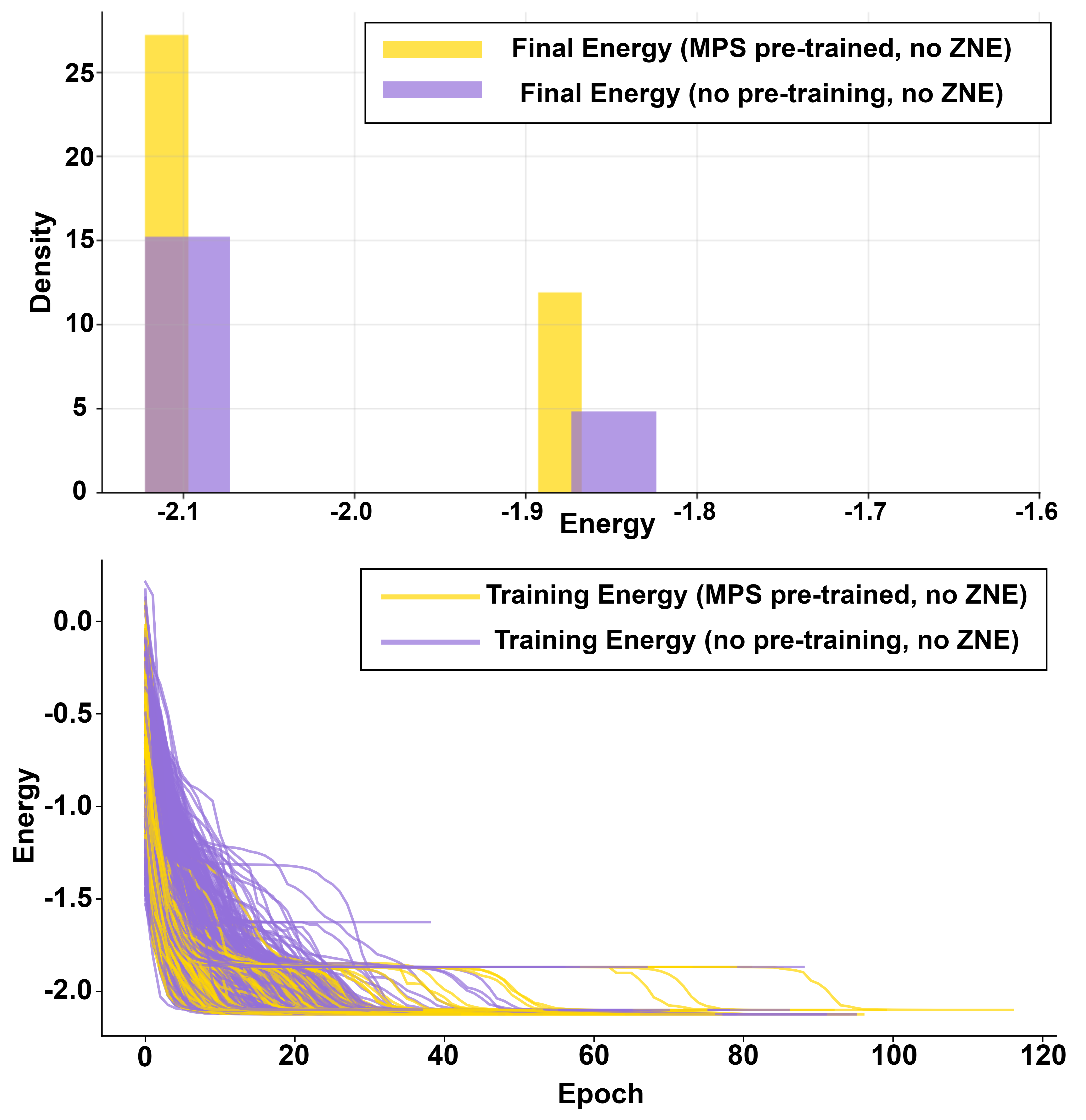} %
    \caption{\label{pre-train}
    Comparison of the ground state energy of the $H_4$ molecule under two conditions: with pre-training (yellow) and without pre-training (purple), based on over 2300 experiments. The upper histogram shows the final energy distributions, where pre-training results in a more concentrated distribution and the dominant bar value closer to the FCI benchmark. The lower plot illustrates the training process, demonstrating better stability and faster convergence with pre-training.
    }
    \end{minipage}
\end{figure}

\textit{Experimental Results.} The Full Configuration Interaction (FCI) benchmark value for this assumed $H_4$ model is -2.1664 Hartree. Table \ref{results} presents the results of different VQE models in solving the ground state energy of the $H_4$ molecule under both ideal noiseless and noisy environments. The listed values are the best results obtained from 30 independent experiments. The comparison results of pre-training MPS to initialize the quantum circuit are shown in Figure \ref{pre-train}. 


Under noiseless conditions, our proposed MPS-VQE circuit, with fewer gates, achieves similar calculation accuracy to the Unitary Coupled Cluster Singles and Doubles (UCCSD) method, which uses approximately 30 times the number of gates.
More importantly, our MPS-VQE demonstrates significantly enhanced noise tolerance compared to other VQE algorithms. 
This indicates that the MPS-VQE can more effectively mitigate the impact of noise interference in actual quantum hardware, thereby exhibiting greater robustness for practical applications.

\section{Conclusion and Discuss}
This work proposes an innovative variational quantum eigensolver that can effectively mitigate the unavoidable noise in NISQ devices. We have introduced innovations in three aspects: circuit initialization, circuit structure design, and error mitigation strategies for both computational and measurement errors. The research findings demonstrate that our proposed MPS-VQE scheme can achieve the precision of large-scale complex circuits using a much more compact circuit, while exhibiting superior noise resilience compared to other VQEs.  

This study provides one new approach for realizing high-precision quantum computing on medium-noise quantum hardware, which is of great significance in advancing the practical applications of quantum computing. Future work may further explore the application of this scheme to more complex quantum chemistry and materials science problems, as well as its generalization to more quantum algorithms, aiming to enhance the practical utility of quantum computation.

\section*{Acknowledgment}   
Thanks for the support provided by MindSpore Community.

\nocite{*}
\bibliography{cites}
\end{document}